\documentclass[10pt]{blois07}
\usepackage{graphicx}
\usepackage{cite,./mcite}

\newcommand{\nc}{\newcommand}
\nc{\del}{\partial}
\nc{\run}[1]{\tilde{\alpha}_{#1}}

\nc{\be}{\begin{equation}}
\nc{\ee}{\end{equation}}
\nc{\bea}{\begin{eqnarray}}
\nc{\eea}{\end{eqnarray}}
\nc{\beast}{\begin{eqnarray*}}
\nc{\eeast}{\end{eqnarray*}}
\def\xBJ{{x_{BJ}}}
\def\xb{{x_{BJ}}}

\def\lsim{\mathrel{\rlap{\lower4pt\hbox{\hskip1pt$\sim$}}
    \raise1pt\hbox{$<$}}}
\def\gsim{\mathrel{\rlap{\lower4pt\hbox{\hskip1pt$\sim$}}
    \raise1pt\hbox{$>$}}}

\begin{document}
\title{What is Measured in Hard Exclusive Electroproduction?}
\author{J.T. Londergan and A.P. Szczepaniak
\protect\footnote{~talk presented at EDS07}}
\institute{Department of Physics and Nuclear Theory Center \\
 Indiana University, Bloomington, IN 47405, USA}
\maketitle

\begin{abstract}
We examine the relation between amplitudes measured in exclusive 
lepto-production and the quark content of the nucleon. We show that in the 
limit of high energy and small $t$, the natural interpretation 
of amplitudes measured in these hard exclusive processes is in terms of 
the quark content of the meson cloud and not the target itself. In this 
regime, Regge amplitudes will make a significant contribution to these 
exclusive amplitudes. This leads to violation of QCD scaling. 
\end{abstract}

\section{Theoretical Analysis of Exclusive Electroproduction}
Recently there has been much interest in hard exclusive reactions. 
This followed from the proof of factorization by Collins \textit{et al.} 
\cite{Collins:1996fb} for exclusive lepto-production. The proof guarantees 
that under certain conditions, the exclusive amplitude can be factorized 
into a hard term calculable from QCD and a soft term that ideally should 
be universal. This latter contribution has typically been parameterized 
by generalized parton distributions or GPDs~\cite{Ji:1998pc,Radyushkin:1997ki,mueller-1988}. This is analogous to the case of deep inelastic scattering (DIS), 
where inclusive cross sections can be parameterized in terms of universal 
parton distribution functions (PDFs) that are related to quark probabilities 
in the nucleon. The only difference is that in hard exclusive processes 
the amplitude typically requires an integral over the GPDs. 

From duality we know that it is in principle possible to use any 
channel to describe the scattering amplitude. In DIS it is known that 
the $s$-channel representation is generally the most efficient way to 
characterize these reactions. The only exception to this occurs at very 
small Bjorken $\xBJ \to 0$. Except in this small-$x$ regime, DIS 
amplitudes can be related to the intrinsic quark structure of the nucleon.  
At very small $x$, amplitudes associated with $t$-channel processes will 
become important; as $\xBJ \to 0$ the structure function evolves to represent 
ladders of partons originating from $t$-channel meson exchanges, and 
Regge exchange makes an important contribution in this regime.  
 
Once it was realized that Regge 
exchange may play a significant role in exclusive electroproduction, attempts 
have been made to incorporate Regge effects using analogies with DIS, 
{\it i.e} to restrict Regge contributions in exclusive electroproduction 
reactions to low-$\xb$ so that scaling is not otherwise modified 
\cite{Ahmad:2006gn,Guzey:2005ec,Kumericki:2007sa,Jenkovszky:2006bq}. 
It had not been proven that Regge contributions 
should only contribute to exclusive amplitudes in this domain. We 
will investigate the question of whether Regge effects should be 
substantial in hard exclusive processes and if so, in what kinematic 
regime they will be important. Rather surprisingly, we find that, at high 
energies when $Q^2$ is large and $t$ small, Regge effects will be significant. 
This implies that in this region, hard exclusive 
processes will be more sensitive to the structure of exchanged mesons 
than they are to the intrinsic quark structure of the nucleon.  

We first investigated this question by examining hard exclusive processes 
in a $t$-channel formalism, as was reported in a recent paper  
\cite{Szczepaniak:2006is}. Consider a general hard exclusive amplitude 
\be
  a^*(q) + N(p) \to b(q') + N(p') \ \ . 
\label{eq:exclus}
\ee
In exclusive electroproduction $a^*(q)$ is a virtual photon with momentum 
$q$, where $-q^2 = Q^2$. In Eq.~(\ref{eq:exclus}), $N(p),N(p')$ represent the 
initial and final nucleons with momenta $p$ and $p'$,  respectively.  
In the Bjorken limit we have $p^2 = p'^2 = m_N^2 << Q^2$, and $b(q')$ denotes 
a final photon or 
meson with momentum $q'$ satisfying $0 \le q'^2 \sim  m_N^2 << Q^2$. As 
is well known, DIS cross sections are proportional to the imaginary part 
of the forward virtual Compton amplitude; this is a special case of  
Eq.~(\ref{eq:exclus}) when $p'=p$ and $q'=q$.  

In Ref.~\cite{Szczepaniak:2006is} we examined the contribution to this 
exclusive amplitude arising from exchange 
of a particle of spin $J$ in the $t$-channel. 
For simplicity we ignored spin and other internal degrees of freedom and 
assumed only scalar currents. The hadronic contribution to 
the cross section is determined from the hadronic tensor, 
\begin{equation}
T(Q^2,\nu,t,q'^2) = \int d^4z e^{i \frac{q + q'}{2} z} \langle p'| T 
  \left[ j(\frac{z}{2}) j(-\frac{z}{2})\right] |p\rangle.
\label{eq:exclusT}  
\end{equation} 
In Eq.~(\ref{eq:exclusT}) $T$ is a function of four independent Lorentz 
scalars with $\nu = p \cdot q/m_N = Q^2/(2\xBJ m_N)$, $t=(p'-p)^2 = 
(q - q')^2$, and $j(z) = \phi^{\dag}(z)\phi(z)$ represents a 
(scalar) quark current in the Heisenberg picture which couples to the 
external fields representing the $a$ and $b$ particles in 
Eq.~(\ref{eq:exclus}).

\begin{figure}
\includegraphics[width=3in]{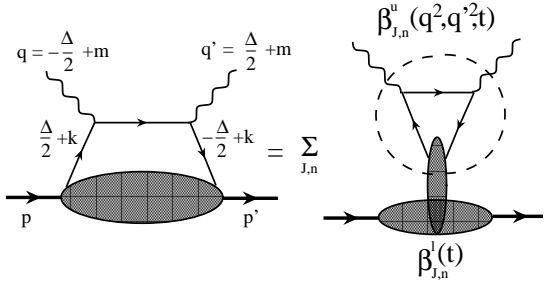}
\caption{\label{Fig:llog}  $t$-channel meson contribution to the hadronic 
tensor for exclusive lepto-production. The amplitude is summed over all spins 
$J$ that can contribute, and depends on the product of the residue functions 
$\beta$ at the upper and lower vertices.}
\end{figure}

In the limit of high energy the contribution to the hadronic tensor from 
$t$-channel exchange of a spin-$J$ meson is proportional to
\bea   
  T_J =  \frac{ \beta^l_{J}(t) \beta^u_{J}(q^2,q'^2,t)}{t - M_J^2}    
 \sum_{\lambda=-J}^J 
&&   \left[ \frac{(p' + p)^{\mu_1}}{2} \cdots \frac{(p'+p)^{\mu_J}}{2} 
  \epsilon^\lambda_{\mu_1 \cdots \mu_J} (p'-p) \right] 
   \nonumber \\ 
 & \times &    \left[\frac{(q' + q)^{\nu_1}}{2} \cdots 
 \frac{(q'+q)^{\nu_J}}{2} \epsilon^{*\lambda}_{\nu_1 \cdots \nu_J} (p'-p) 
 \right]. \nonumber \\
\label{eq:TJ}
\eea 
 In Eq.~(\ref{eq:TJ}), $\epsilon$ is the spin-$J$ polarization vector, and 
$\beta^l_{J}$ and $\beta^u_{J}$ are the residue functions at the {\it lower} 
and {\it upper} vertex, respectively. This is shown schematically in 
Fig.~\ref{Fig:llog}. In the Bjorken limit, 
 $s \to Q^2(1-\xBJ)/\xBJ$ and the amplitude reduces to 
 \be 
  T_J =  \frac{\beta^l_{J}(t) \beta^u_{J}(q^2,q'^2,t)}{t - M_J^2}   
  \left( \frac{Q^2}{2\xBJ} \right)^J .
\label{eq:TJred}
\ee

The key question is how the upper residue function depends on the large 
variables ($Q^2$ and $-q'^2 = Q^2$ in the case of inclusive processes, and 
$Q^2$ for the exclusive amplitudes). It is well known that for kinematics 
relevant to inclusive scattering, the upper residue function 
behaves as $(1/Q^2)^{J+1}$, {\it modulo} logarithmic corrections, so that the  
amplitude scales,  $ Q^2 T_J  \propto  (1/\xBJ)^{J}$, as 
expected~\cite{Ioffe,Brodsky:1971zh,Brodsky:1972vv,Brodsky:1973hm}. 
Summing over all spins leads to the Regge behavior, $Q^2 T = \sum_J T_J 
\propto (1/\xBJ)^{\alpha(0)}$.  In the small $\xBJ \to 0$ limit in DIS, 
the leading Regge trajectory with 
$\alpha(0)>0$ will dominate the behavior of the hadronic tensor, 
while all daughter trajectories with $\alpha_n(0) < \alpha(0)$ are subleading 
at very small $x$.  For finite $\xBJ$, however, daughter Regge trajectories 
are no longer suppressed, and as a result the Regge description becomes 
ineffective while the $s$-channel parton model description becomes natural.

The situation is quite different for exclusive electroproduction. This  
can be shown by writing the contribution of a $t$-channel spin $J$ exchange 
in terms of the two-current correlation in the exchanged meson. 
This can be expanded in terms of a covariant Bethe-Salpeter amplitude 
for the exchanged meson, represented in terms of the spectral 
density $g_{Jn}$. The spectral density is related to the parton distribution 
amplitude in a meson and can in principle be constrained from electromagnetic 
data~\cite{Szczepaniak:1993kk} and QCD asymptotics~\cite{Lepage:1980fj}. 
In the Bjorken limit, using the Feynman parametrization for the quark 
propagators and ignoring small terms we showed that the upper vertex function 
could be written \cite{Szczepaniak:2006is}
\bea
\beta^u_{Jn}  &=&  \int_{-1}^1 dx \int d\mu^2  g_{Jn}(x,\mu^2) \int_0^1 
d\alpha \frac{\alpha^J }{\left[ -\alpha \left(  \frac{q'^2 + q^2}{2}  
 + x\frac{ q'^2 - q^2}{2}\right) + \mu^2\right]^{n+J-1}  } \ .
\nonumber \\ 
\label{eq:beta} 
\eea
For inclusive amplitudes when $q'^2=q^2=-Q^2$, the $x$ disappears from the 
denominator and the integration over $\alpha$ is 
dominated by $\alpha\sim \mu^2/Q^2$.  As a result, the entire integral is of 
order $(\mu^2/Q^2)^{J+1}$, as we argued earlier. 

For exclusive amplitudes however where $q'^2 \sim 0$, the integrand of 
Eq.~(\ref{eq:beta}) is dominated by the region 
$1-x = {\mathcal O}(\mu^2/Q^2)$, and finite $\alpha$. The endpoint behavior of 
the distribution amplitudes $g_{nJ}$ is spin independent, and for 
leading-twist amplitudes  $g_{Jn}(x \to 1) \sim (1-x)$.  This leads to 
a $J$-independent suppression of the upper vertex with $Q^2$, $\beta^u_{Jn} 
\sim {\mathcal O}(\mu^4/Q^4)$ {\it independent of the spin of the 
exchanged meson}. Upon summing over all spins from a single trajectory one 
finds that for small $t$ the hadronic tensor is proportional to 
$(Q^2/\xBJ)^{\alpha(t)}$. Thus, in the Bjorken limit exclusive 
lepto-production should be dominated by a single leading Regge trajectory 
for all $\xBJ$, and not just for $\xBJ \to 0$.  In this regime hard 
exclusive processes probe the nucleon's meson cloud rather than its intrinsic 
quark properties.  

Since this result was rather surprising, we repeated the derivation of hard 
exclusive amplitudes in an $s$-channel framework \cite{Szc07}, where we 
analyzed the ``handbag'' diagrams used in extracting GPDs.  We obtained the 
same results, namely that in the region of high energy and small $t$, Regge 
effects will make sizeable contributions to hard exclusive amplitudes. We 
showed that the DVCS formalism is ill-defined in the presence of Regge 
behavior in the parton-nucleon amplitude. This has the following consequences 
for hard exclusive processes.  
\begin{enumerate}
\item The breakdown of collinear factorization in these processes means 
that the soft amplitudes are not universal, but are process-dependent 
amplitudes that we call Regge Exclusive Amplitudes. 
\item The validity of QCD factorization for exclusive lepto-production will 
require, in addition to a hard scale $Q^2$, a sizeable value of $t$.  
\item In the region of small $t$ Regge effects will make substantial 
contributions to DVCS and exclusive meson lepto-production. 
\item Exclusive lepto-production processes will show violation of QCD scaling 
arguments. These scaling violations will persist regardless of the magnitude 
of $Q^2$, in contradiction to expectations of QCD scaling. 
\item In the region of small $t$ hard exclusive amplitudes will 
exhibit a $Q^2$ behavior $(Q^2/\xBJ)^{\alpha(t)}$ characteristic of hadronic 
Regge amplitudes. These amplitudes should be well approximated by the 
contribution from the leading Regge trajectory. 
\end{enumerate}

\section{Experimental Support for Regge Exchange}  

Recently a Hall A Collaboration at Jefferson Laboratory has carried out 
a test of scaling in DVCS reactions \cite{MunozCamacho:2006hx}. The 
data appear to be in very good agreement with the  
$Q^2$-independent DVCS amplitude predicted by QCD \cite{Vanderhaeghen:1998uc}, 
however the available $Q^2$ window is quite small, from 
$1.5 - 2.5\,\mbox{GeV}^2$ and within the published experimental errors one 
cannot rule out a power-like dependence of the amplitude, $A \propto 
(Q^2)^\alpha$, with $\alpha$ as large as 0.25.  Even more surprising, 
"standard" Regge-exchange models have proved successful in describing a 
variety of differential cross sections \cite{Morand:2005ex,Kubarovsky}, in the 
kinematic range where scaling would be expected based on comparisons with 
DIS.  Our calculations show that the success of the 
Regge picture is to be expected, and is not accidental.  

\begin{figure}
\includegraphics[width=2.5in,angle=270]{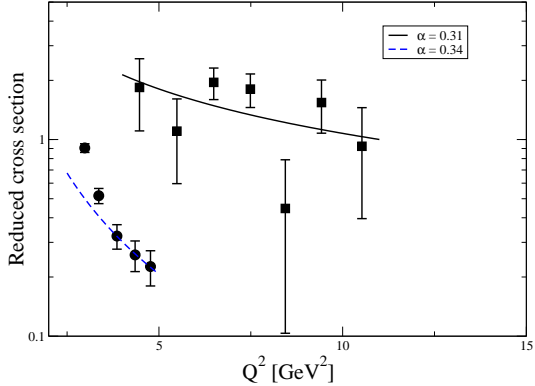}
\caption{\label{datos2} A simple fit to some meson electroproduction data;  
$\pi^+$ results from HERMES 
(squares,~\protect\cite{HERMES-from-jlab}), and $\omega$ 
(circles,~\protect\cite{Morand:2005ex}) results from the CLAS Collaboration at 
Jefferson Lab. In the case of $\pi^+$ production the cross section reduced by 
the photon flux is plotted in arbitrary units.}  
\end{figure}

A recent experimental analysis of $\omega$ 
electroproduction at Jefferson Laboratory~\cite{Morand:2005ex} 
showed that their data was in good agreement with predictions from standard 
Regge phenomenology, while showing large uncertainties with analyses based on 
models of GPDs~\cite{Diehl:2005gn}. Though our results 
were derived in the Bjorken limit with $s/|t| >> 1$, the JLab data  
corresponds to energies of a few GeV and values up to $|t| \sim 2.7$.  
In Fig.~\ref{datos2} we compare our predictions with   
exclusive meson electroproduction data. QCD scaling arguments predict that the 
reduced $\pi^+$ cross section should fall off at fixed $\xb$ as $1/Q^2$.  
We predict a behavior $(Q^2)^{2\alpha-1}$ with $0<\alpha  < 1$. Fitting 
$\pi^+$ data from HERMES~\cite{HERMES-from-jlab} in the range  
$0.26 < \xb < 0.8$ gives $\alpha =0.31 \pm 0.2$. Similarly for $\omega$ 
electroproduction cross section from the CLAS 
collaboration~\cite{Morand:2005ex} we find $\alpha =0.34 \pm 0.24$  
for the range $0.52 < \xb < 0.58$. 

We see that for both DVCS and exclusive meson electroproduction, not only are 
scaling violations observed, but the additional $Q^2$ dependence is softer 
than predicted by scaling, and is in 
agreement with our predicted factor of $(Q^2)^{\alpha}$ where  
$0< \alpha < 1$.  At this point it is difficult to compare the Regge exponents 
$\alpha$ obtained from the fit with total cross-section data, since the 
electroproduction data corresponds to different values of $t$. This issue 
warrants further phenomenological study. 

In conclusion, we have shown that the Regge nature of parton-nucleon  
amplitudes generates divergences in GPDs at low $t$. In this region we 
predict sizeable effects due to scattering from the meson cloud in the 
nucleon, while at sufficiently large $t$ the dominant effect will come 
from scattering of quarks in the nucleon. An important remaining question is 
how one can disentangle scattering off 
the meson cloud from effects of nucleon tomography.

Some of the work here was done in collaboration with F. Llanes-Estrada. The 
authors would like to thank S. Brodsky, J-M. Laget, W. Melnitchouk, 
D. Mueller, A. Radyushkin, M. Strikman and C. Weiss for useful comments 
and discussions. JTL was supported by NSF contract PHY-0555232 and APS by 
DOE contract DE-FG0287ER40365.

%------------------------------------------------------------------------------
%       Bibliography
%------------------------------------------------------------------------------
\begin{footnotesize}
\bibliographystyle{blois07} 
{\raggedright
\bibliography{londergan-eds07}
}
\end{footnotesize}

\end{document}